\documentclass{cimento}

\usepackage{graphicx}  
\usepackage{fancyhdr}
\title{Higgs sector of the MSSM: lepton flavor violation
at colliders and neutralino dark matter\thanks{Prepared for the proceedings of the workshop: "LC09:
$e^+ e^-$ Physics at the TeV Scale and the Dark Matter Connection", 21-24 September 2009, Perugia, Italy }}

\author{M.~Cannoni\from{ins1} \atque O.~Panella\from{ins2}}

\instlist{
\inst{ins1} Universidad de Huelva, Departamento de Fisica Aplicada, Campus de El Carmen - 21071 Huelva, Spain
\inst{ins2} Istituto Nazionale di Fisica Nucleare, Sezione di Perugia, Via A.~Pascoli, 06129, Perugia, Italy
}

\PACSes{
\PACSit{11.30.Pb} {Supersymmetry}
\PACSit{14.80.Da} {Supersymmetric Higgs bosons}
\PACSit{95.35.+d} {Dark matter (stellar, interstellar, galactic, and cosmological)}
}

\begin{document}

\maketitle

\begin{abstract}
We examine the prospects for the detection of Higgs mediated lepton flavor violation at
LHC and at a photon collider in the minimal supersymmetric standard model with 
large lepton flavor violating mass insertions in the $\mu-\tau$ sector
constraining  the parameter space with several experimental bounds.
We find rates
probably too small to be observed at future experiments if models have to 
accommodate for 
a neutralino relic density as measured by WMAP and explain the $(g-2)_{\mu}$ anomaly:
better prospects are found if these two constraints are applied
only as upper bounds.  
The spin-independent
neutralino-nucleon cross section in the studied constrained parameter space
is just below the present CDMS limit while gamma rates from neutralino annihilation
in the halo are strongly suppressed.

\end{abstract}

\section{Introduction}

The Higgs sector of the minimal supersymmetric standard model (MSSM)~\cite{djouadi}, especially the 
heavy neutral Higgses $A$ and $H$, play a prominent role in the physics
of neutralino dark matter~\cite{kamionkowski}.
In some region of the supersymmetric (SUSY) parameter space
neutralinos yield the desired amount of relic density by annihilating 
into fermions through the $s$-channel resonant exchange of neutral Higgs bosons $h$, $H$, $A$,
the so called funnel region where $m_A \simeq 2 m_{\chi}$; besides, 
as dark matter is expected to be distributed 
in a halo surrounding our galaxy,  
neutralinos can scatter off nuclei in terrestrial detectors: the coherent scattering is mediated
by scalar interactions through the $s$-channel exchange of squarks and 
$t$-channel exchange of the CP-even neutral 
Higgs bosons $h$ and $H$. 
These effects become sizable when squarks are heavy 
and $\tan\beta$ is large in reason of the enhanced Higgs bosons coupling to
down-type fermions, especially for the $b$ quark which has the
largest Yukawa coupling receiving 
large radiative SUSY-QCD corrections at large $\tan\beta$.
  
Once a source of lepton flavor violation (LFV) is present in the slepton mass matrix, for example
the MSSM with the see saw mechanism for generation of small 
neutrino masses~\cite{barger1},
non-holomorphic LFV Yukawa couplings of the type $\bar{L}_R^i L_L^j H_u^*$ are induced at loop
level and become particularly sizable at large $\tan\beta$ giving rise to enhanced 
Higgs-mediated LFV effects~\cite{bkl}.
The LFV mass insertions 
$\delta^{ij}_{LL}\!=\!{({m}^2_{L})^{ij}}/{m^{2}_{L}}$,
$\delta^{ij}_{RR}\!=\!{({m}^2_{R})^{ij}}/{m^{2}_{R}}$,
where $({m}^2_{L,R})^{ij}$ are the off-diagonal flavor changing entries
of the slepton mass matrix, are free parameters which allow for
a model independent study of LFV signals. 
We introduce LFV in the model through the  
mass insertions $\delta_{LL,RR}^{32} =0.5$.
This value ensures the largest rates in LFV
processes and allow us to study the more optimistic scenarios for LFV detection; 
higher values 
contradict the mass insertion approximation as an expansion of propagators in
this small parameters.

Higgs mediated effects become interesting at large $\mu$ and $\tan\beta$ and low $m_A$;
further, if SUSY-QCD particles are heavy, Higgs effects are dominant also for neutralino 
dark matter.
We thus consider the following real MSSM parameter space:
100 GeV $\le m_A \le $1 TeV,
20$\le\tan\beta\le$60,
500 GeV$\le \mu \le$5 TeV,
(the sign of $\mu$ is taken positive, as preferred by 
the SUSY explanation of the $(g-2)_\mu $ anomaly), 
150 GeV$\le M_1, M_2\le $1.5 TeV
(we do not impose any relation
but let them vary independently allowing for gaugino non-universality at the weak scale), 
1 TeV $\le M_3\le$ 5 TeV (to have large masses for gluinos),
1 TeV$\le m_{U_3}, m_{D_3}, m_{Q_3} \le$5 TeV
(for the first and the second generation the soft masses
are set to be equal, $ m_{U_i} = m_{D_i}
= m_{Q_i} =m_{\tilde{q}}$,
where $i=1,2$ and $m_{\tilde{q}}$ is another free parameter
which varies in the same range), 
300 GeV$\le m_{L_3}, m_{E_3} \le $ 2.5 TeV
(for the first and the second generation the slepton soft masses
are set to be equal, $ m_{L_i} = m_{E_i}
=m_{\tilde{\ell}}$, 
where $i=1,2$ and $m_{\tilde{\ell}}$),
-2$\le \frac{A_{U_3}}{m_{U_3}}$, $\frac{A_{D_3}}{m_{D_3}}$, $\frac{A_{E_3}}{m_{E_3}} \le 2$
while for first and second generation the trilinear scalar
couplings are set to zero. 

We impose on the parameter space several experimental limits:
(1) LEP, TEVATRON bounds on sparticle masses and
the LEP bound on light Higgs;
(2) present bounds on  $B$-physics observables 
${\cal B}(B \to X_s \gamma)$, ${\cal B}(B_s \to \mu^+ \mu^-)$, ${\Delta m_{B_s}}$, ${{\cal B}(B\to\tau\nu)}$;
(3) the present experimental upper bounds on LFV processes
$\mathcal{B}(\tau\to\mu\gamma)$,
$\mathcal{B}(\tau\to\mu\eta)$,
$\mathcal{B}(\tau\to\mu\mu\mu)$,
(4) the exclusion limits on the neutralino-nucleus spin independent cross section
from the CDMS experiment~\cite{cdms};  
(5) the CDF exclusion exclusion limits in the ($m_A$, $\tan\beta$) plane~\cite{cdf}.
All of them are applied at 
the same time with the exception of the WMAP $3\sigma$ interval on relic density 
and $(g-2)_\mu$ anomaly 
for which we also relax the lower bounds: thus in the following figures
the light gray (turquoise) points 
have only $\Omega_{\chi} h^2 \le 0.13$ and $a^{MSSM}_{\mu} \le 4\times 10^{-9}$, 
the plus-shaped points dark-gray (indigo) points
satisfies $0.09 \le \Omega_{\chi} h^2 \le 0.13$ and $a^{MSSM}_{\mu} \le 4\times 10^{-9}$,  
finally, the squared points satisfy  $0.09 \le \Omega_{\chi} h^2 \le 0.13$ and  
$1\times 10^{-9} \le a^{MSSM}_{\mu} \le 4\times 10^{-9}$.
For numerical computations we use 
the code \textsc{DarkSusy}~\cite{darksusy} and the code \textsc{FeynHiggs}~\cite{feynhiggs},
while for the explicit formulas and further details on the experimental constraints we refer the reader to 
ref.~\cite{cannoni1}.

\section{Higgs mediated LFV at LHC and at a photon collider}
\begin{figure}[t!]
\begin{center}
\includegraphics*[scale=0.5]{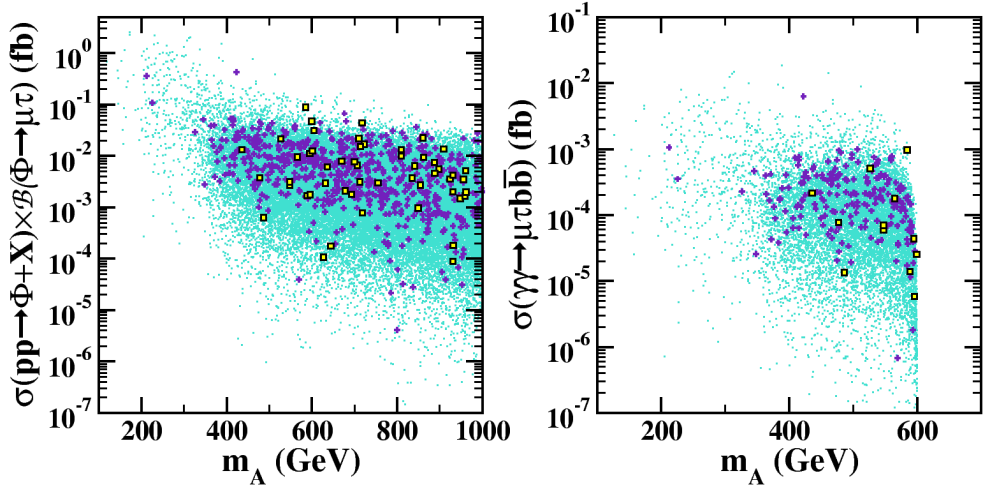}
\caption{{ Left:} Scatter plot of the inclusive production cross  
section $pp\to\Phi +X$ times the branching ratio of $\Phi\to\tau\mu$ at LHC 
versus $m_A$ ($\Phi =A, \,H$). 
{Right:} Scatter plot for the cross section of the process $\gamma\gamma\to\tau\mu b\bar{b}$
in photon-photon collision at $\sqrt{s_{\gamma\gamma}} = 600$ GeV. 
The light gray (turquoise) points 
have only $\Omega_{\chi} h^2 \le 0.13$ and $a^{MSSM}_{\mu} \le 4\times 10^{-9}$, 
the plus-shaped points dark-gray (indigo) points
satisfies $0.09 \le \Omega_{\chi} h^2 \le 0.13$ and $a^{MSSM}_{\mu} \le 4\times 10^{-9}$,  
finally, the squared points satisfy  $0.09 \le \Omega_{\chi} h^2 \le 0.13$ and  
$1\times 10^{-9} \le a^{MSSM}_{\mu} \le 4\times 10^{-9}$.}
\label{fig1}
\end{center}
\end{figure}
At high $\tan\beta$ the dominant production mechanisms
for $A,H$ at LHC is $b\bar{b}$ fusion
due to the $m_b \tan\beta$ enhanced $b\bar{b}\Phi$ couplings.
We calculate the cross section with \textsc{FeynHiggs} 
which uses the approximation  
%\beq
$\sigma^{MSSM}(b\bar{b}\to \Phi) = \sigma^{SM} (b\bar{b}\to \Phi ) 
\frac{\Gamma(\Phi \to b\bar{b})^{MSSM}}{\Gamma(\Phi \to b\bar{b})^{SM}}$,
%&\eeq
where $\sigma^{SM}(b\bar{b}\to \Phi)$ is the total SM cross section
for production of Higgs boson with mass $m_{\Phi}$ via $b\bar{b}$ fusion: to obtain the 
value in the 
MSSM it is rescaled with the ratio of the decay width of the inverse process in
the MSSM over the SM decay width~\cite{feynhiggs}.
We calculate for each random model the product the
$\sigma(pp\to \Phi +X)\times{\cal B}(\Phi\to\tau\mu)$.
As masses and couplings of $A$ and $H$ 
are practically identical as discussed above, we have 
$\sigma(pp\to A+X) + \sigma(pp\to H+X)\simeq 2 \sigma(pp\to A+X)$. 
The scatter plot 
$\sigma(pp\to \Phi +X)\times{\cal B}(\Phi\to\tau\mu)$ is shown in Fig.~\ref{fig1}, left panel. 
We see that with 
the nominal integrated luminosity of $100$ fb$^{-1}$ per year models 
which satisfy both the relic density abundance and $\Delta a_\mu$ 
can give up to 10 events per year (squared points), up to 40 if we relax the condition
on the lower limit of $\Delta a_\mu$ (plus-shaped points) and up to 200-300 relaxing
both the lower limits (turquoise (light-gray)  points).

In $\gamma\gamma$ collisions the main production mechanism for $\Phi =A,H$
is $\tau\tau$ fusion while the $b\bar{b}$ is suppressed
by a factor $3(1/3)^4 (m_b /m_{\tau})^2 \simeq 0.1$ which cannot
be compensated by corrections to the $b$ Yukawa coupling.
In ref.~\cite{cannoni} we studied in detail
the $\mu\tau$ fusion process $\gamma\gamma \to \mu\tau b\bar{b}$
where the Higgs boson is produced in the $s$-channel via a virtual $\mu\tau$ pair and
can be detected from its decay mode $A \to b\bar{b}$.  
There we have shown that a good analytical approximation for the cross section is obtained using
the equivalent particle approximation wherein the colliding
real photons split respectively into $\tau$ and $\mu$ pairs with the subsequent
$\mu\tau$ fusion into the Higgs boson and that the effect of photons spectra 
can be neglected. 
we thus consider  
monochromatic photons with $\sqrt{s_{\gamma\gamma}} =600 $ GeV,
and photon-photon luminosity 
500 fb$^{-1}$ yr$^{-1}$~\cite{cannoni}.
The scatter plot of the signal cross section versus $m_A$ is shown in Fig.~\ref{fig1},
right panel. Here the models 
which satisfy both the relic density abundance and $\Delta a_\mu$ (squared points)
have maximal cross section $10^{-3}$ fb, which is too small.
Relaxing the lower limits cross section values up to $2\times 10^{-2}$ fb are possible,
giving 10 events/year.

\section{Neutralino dark matter direct and indirect detection}
\begin{figure}[t!]
\begin{center}
\includegraphics*[scale=0.5]{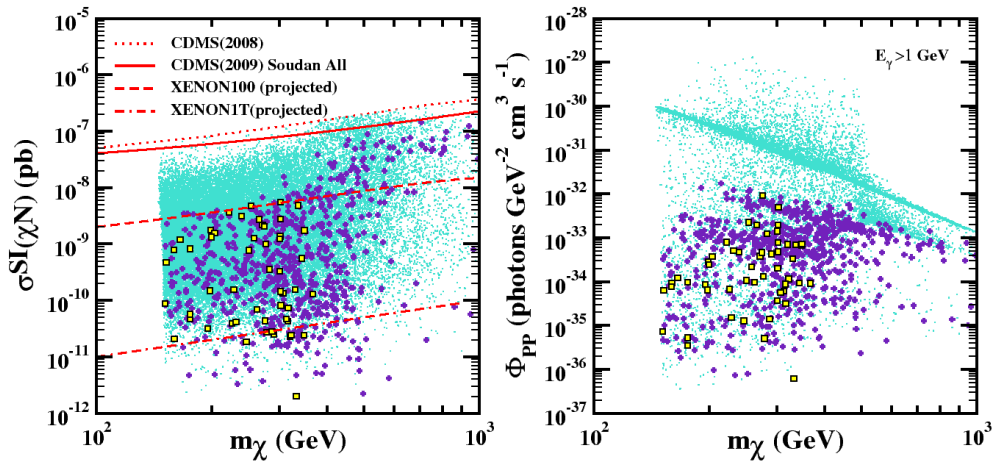}
\caption{ 
{Left:} Scatter plot for the spin-independent neutralino-nucleon
cross section versus the neutralino mass. The area above the solid line is excluded 
the CDMS final results; the area above the dotted line is excluded by the 2008 CDMS
search.
The dashed and dot-dashed lines give the sensitivity reach 
of two phases of the XENON experiment. 
{Right:} Scatter plot for the particle physics factor entering the formula 
of the flux of gamma rays from neutralino annihilation in the halo, see the text.
The light gray (turquoise) points 
have only $\Omega_{\chi} h^2 \le 0.13$ and $a^{MSSM}_{\mu} \le 4\times 10^{-9}$, 
the plus-shaped points dark-gray (indigo) points
satisfies $0.09 \le \Omega_{\chi} h^2 \le 0.13$ and $a^{MSSM}_{\mu} \le 4\times 10^{-9}$,  
finally, the squared points satisfy  $0.09 \le \Omega_{\chi} h^2 \le 0.13$ and  
$1\times 10^{-9} \le a^{MSSM}_{\mu} \le 4\times 10^{-9}$.}
\label{fig2}
\end{center}
\end{figure}
The spin-independent neutralino-nucleon cross section 
in the limit of heavy squarks and large $\tan\beta$ 
can be approximated as~\cite{carena1}  
%\beq
$\sigma^{SI}\simeq \frac{{g'}^2 g^2 |N_{11}|^2 |N_{13}|^2 m_N^4 }{4\pi m_W^2 m_A^4}\tan^2 \beta\times K_{f}$,
%\label{sif}
%\eeq
where $N_{11}$ and $N_{13}$ are the lightest neutralino
unitary mixing matrix elements, $m_N$ the nucleon mass (neglecting the mass difference
between the neutron and the proton) and $K_{f} $
a factor which depends  on nucleon form factors.
The left panel of Fig.~\ref{fig2} shows the scatter plot for the spin-independent 
neutralino-nucleon cross section as a function of $m_\chi$ and the 
region excluded by CDMS~\cite{cdms}.
We emphasize that 
CDF and CDMS limits are very mild constraints: 
the region excluded by CDF is practically excluded by the other constraints
while the CDMS limit exclude only one plus-shaped point 
leaving untouched the regions preferred by WMAP and the $(g-2)_\mu$ anomaly.
The XENON100 experiment~\cite{xenon} should 
reach the sensitivity corresponding to the dashed gray (red) line in the Figure~\ref{fig1}, left panel.
Such sensitivity is able to cover the region with the highest cross 
section, $m_\chi \ge 300$ GeV, where there is large higgsino component.
On the other hand
the region preferred by $(g-2)_\mu$ anomaly cannot be covered.
We also report the prospected sensitivity goal of the XENON experiment 
with 1 ton detector mass~\cite{xenon}, dot-dashed gray (red) line, which is 
$10^{-11}-10^{-10}$ pb for neutralino mass in the range 
$100-1000$ GeV: practically all of the parameter space 
can be probed. 

We also present the effect of our constrained parameter space 
on the flux of photons coming from neutralino annihilation in the halo which is
a very active field in indirect dark matter detection. The flux of gammas
expected from neutralino annihilation is generally given by 
$F = \Phi_{PP} \times \Phi_{astro}$ where the second factor contains the astrophysical 
informations and is given by the integral of the squared of the dark matter density
along the direction of observation, while  $\Phi_{PP}$ is the particle physics factor
given by 
$\Phi_{PP} (E_\gamma > E_{th}) =  \langle \sigma v \rangle/(2 m_{\chi}^2) \int_{E_{th}}^{m_\chi} dE (dN_{\gamma}/dE)$
where $\langle \sigma v \rangle$ is the thermal averaged cross section annihilation times the velocity
of neutralinos and the $dN_{\gamma}/dE$ is photon spectrum which is integrated over energies greater than 
$E_{th}$. Here we are interested mainly in the evaluation of the particle physics factor and in comparing 
it with other other studies in literature in the framework of the MSSM.  
We present  $\Phi_{PP}$ as a function of the neutralino mass in fig.~\ref{fig2}, right panel,
with threshold energy $E_{th} =1$ GeV. 
For models with a relic density inside the WMAP interval the maximum value of the particle physics 
factor is $2\times 10^{-32}$ photons GeV$^{-2}$ cm$^3$ s$^{-1}$ for neutralino mass around $200$ GeV:
this value is two order of magnitude smaller than the typical values found in similar studies 
without the constraints from lepton flavor violation and the updated $B$ physics and $(g-2)_\mu$ 
constraints, see refs.~\cite{fornengo}.     
We remark that most of the models which satisfy the WMAP bounds, the plus-shaped and squared
points in the figures, satisfy the Higgs funnel condition $m_A \simeq 2 m_{\chi}$ (see fig.~2, left panel)
of ref.~\cite{cannoni1}): the main annihilation channel is $\chi\chi \to b\bar{b} / \tau\bar{\tau}$ through $s$-channel 
heavy Higgs exchange which is strongly constrained in our scenario and thus it is natural to 
have a reduction of photons emission.  A reduction of the gamma ray flux in the funnel region it also
found in ref.~\cite{barger1} in the study of the minimal supergravity (mSUGRA) plus right-handed neutrinos
for see-saw generation of neutrino masses respect to the case of pure mSUGRA.

\section{Summary and conclusions}

In the framework of the MSSM with heavy SUSY-QCD
particles and large $\tan\beta$ 
we have studied lepton flavor violation in $\tau-\mu$ sector mediated by
the heavy neutral Higgs $\Phi=A$, $H$
at high energy colliders through the production 
and decay at LHC, $p p \to \Phi +X$, $\Phi \to \tau\mu$ and the $\mu-\tau$ fusion at 
a photon collider, $\gamma\gamma\to\tau\mu b\bar{b}$. 

We have found that in models with $0.09 \le \Omega_{\chi} h^2 \le 0.13$ 
and $1\le a_{\mu}^{MSSM}\times 10^{9} \le 4$: 
at LHC the cross section for   
$p p \to \Phi +X$, $\Phi \to \tau\mu$ can reach ${\cal O}(10^{-1}-10^{-2})$ fb in the
range $m_A =400-1000$ GeV giving up to 10 events with 100 fb$^{-1}$; 
the cross section of $\gamma\gamma\to\tau\mu b\bar{b}$ reaches ${\cal O}(10^{-3})$ fb,
thus too small even for the large value of the expected luminosity of 500 fb$^{-1}$.
Prospects are somewhat more encouraging if we relax the lower limits, imposing only
$\Omega_{\chi} h^2 \le 0.13$ and $a_{\mu}^{MSSM}\times 10^{9} \le 4$: 
the cross section at LHC is about 2 fb for low $m_A$ masses and around $2\times 10^{-2}$ fb
in $\gamma\gamma$ collisions. 
On the other hand, to observe such effects, in any case, the full luminosity of the machine is needed.

We have also studied the spin-independent neutralino nucleus cross section: we have shown that   
in models that satisfy  $0.09 \le \Omega_{\chi} h^2 \le 0.13$ and $1\le a_{\mu}^{MSSM}\times 10^{9} \le 4$,
the cross section lies just below the sensitivity of XENON100 which should report 
results soon. The full XENON 1 ton is needed to cover all the parameter
space. 
For models with a relic density inside the WMAP interval the particle physics 
factor for gamma rays flux from neuralino annihilation in the haloes is found to be 
smaller than the typical values found in similar studies 
without the constraints from lepton flavor violation and the updated $B$ physics and $(g-2)_\mu$ 
constraints.

\acknowledgments
M.~C. acknowledges support by the project
P07FQM02962 funded by "Junta de Andalucia", partial support
by the FPA2008-04073-E/INFN project and  
MULTIDARK project of Spanish Ministry of Science and Innovation's  Consolider-Ingenio  
Ref: CSD2009-00064.

\end{document}